\documentclass[12pt]{article}
\usepackage{fullpage}
\usepackage{graphicx}
\usepackage[nooneline,tight]{subfigure}\subfigcapskip=-12pt
\usepackage{hyperref}
\usepackage{natbib}
\def\gesim{\ \hbox to 0 pt{\raise .6ex\hbox{$>$}\hss}\lower.5ex\hbox{$\sim$}\ }
\def\lesim{\ \hbox to 0 pt{\raise .6ex\hbox{$<$}\hss}\lower.5ex\hbox{$\sim$}\ }
\begin{document}
\title{Intergrain forces in low Mach-number plasma wakes}
\author{I H Hutchinson}
\date{Plasma Science and Fusion Center and\\ Department of Nuclear
  Science and Engineering,\\ Massachusetts Institute of Technology,
  Cambridge, MA, USA}


\maketitle

\begin{abstract}
Large-scale particle-in-cell calculations of the plasma wake
interactions of two negatively charged grains smaller than the Debye
length are carried out using the COPTIC code over a wide range of
subsonic plasma flow velocities. In plasmas with temperature ratio
$T_e/T_i=100$, it is found that a single grain's oscillatory wake
disappears for flow Mach numbers ($M$) less than approximately 0.3,
which is the parameter regime where Landau damping is expected to be
strong. Neutral collisions suppress potential oscillations above
$M=0.3$, but not the trailing attractive potential peak caused by ion
focussing. The transverse (grain-aligning) force on a downstream
particle in the wake of another is obtained rigorously from the code
in three-dimensional simulations. It shows general agreement with the
force that would be deduced from the single-grain wake potential
gradient. Except for relatively large grains in the nonlinear
collisional regime, the grain-aligning force is very small for slow
flow.
\end{abstract}

\section{Introduction}

The wake formed by plasma flow past a charged object is a phenomenon
important for solar-system plasma physics, space-craft interactions,
and for understanding Langmuir probes. It also controls the effective
inter-grain interaction in dusty plasmas; and so the objects embedded in
the plasma will for consistency be called ``grains'' in this paper.  The
force between grains in flowing plasmas is known to be anisotropic and
non-reciprocal. This character arises from the effect of the wake of
one upon the other. It is widely found, for example, that grains
suspended by the balance of forces in a plasma sheath edge often align
themselves vertically\cite{chu94,melzer96,takahashi98,melzer99,steinberg01,hebner03,hebner04,kroll10}
 because the downstream grain experiences an
attraction towards the flow axis of the upstream grain's wake. The
attraction arises from focussing of ions (attracted by the negatively
charged grain) enhancing the ion density in the immediate downstream
region. The flow velocity in these situations is approximately equal
to the sound speed.

Linear response formalism\cite{rostoker61} has long been used to
calculate the anticipated form of the wake potential of a single point
charge (grain)\cite{sanmartin71,chen73,lampe00}. Dust grains in many
experiments have large enough charge, however, that they are not well
represented by a linearized calculation. Early non-linear calculations
based upon fluid plasma representations
(e.g.\ \cite{Stangeby1971,Melandso1995}) omit kinetic phenomena, such
as ion orbit-crossing and Landau damping, that are crucial to the
physics. Recent kinetic collisionless particle-in-cell
calculations\cite{hutchinson11} agree with linear response
calculations in the linear regime, but have shown that the
non-linearities substantially suppress the wake amplitude at
experimental parameters. They also have confirmed\cite{Hutchinson2011}
(contradicting some prior claims) that the intergrain force for sonic
flow is fairly well represented by the gradient of the single-grain
wake potential acting on the charge of the downstream grain.

As the flow velocity is reduced below the sound speed, the oscillatory
wake does not immediately disappear. Nevertheless, it must eventually disappear and revert to
the spherically symmetric behavior of a zero-flow situation. The
question then arises as to where and how this transition takes place.  

Experiments on dusty plasmas in micro-gravity
conditions (e.g.\cite{Annaratone2002}) have shown that there appears to be a
mechanism aligning chains of grains far from the sheath. And
recent analysis indicates this alignment occurs in situations where the
ion flow velocity is much smaller (by a factor of ten or more) than
the sound speed\cite{arp12}. These observations give extra incentive
to explore the behavior of the wake in low Mach-number flows. That is
the purpose of the present computational investigations.

\section{Code Description}

The present results are obtained with the Cartesian mesh, Oblique
boundary, Particles and Thermals in Cell (COPTIC) code. It is a hybrid
PIC code in which the electron density is presumed governed a simple
thermal Boltzmann factor $n_e=n_{e\infty}\exp(\phi e/T_e)$. The ions
are represented by particles moving according Newton's 2nd law under
the influence of the self-consistently calculated electrostatic
potential $\phi$. Objects of various shapes can be embedded in the
three-dimensional cartesian computational mesh, and their boundaries
are treated with second order accuracy by the finite difference scheme
even if their surfaces are oblique to the mesh\cite{hutchinsona11}. In
addition, infinitesimal point-charge grains can be included using a
PPPM\cite{hockney88} scheme. And in the present work, only point
grains are included. Prior studies have shown that such calculations
are fully representative of grains with radius up to one tenth of the
Debye length ($r_p\le 0.1 \lambda_{De}$ where $\lambda_{De}^2=
\epsilon_0 T_e/n_e e^2$). Fuller description of COPTIC has already
been published\cite{hutchinson11}.

In the present work, calculations have been performed with and without
charge-exchange collisions. The collision implementation is the same
as for the SCEPTIC code\cite{Hutchinson2007a}, consisting of
Poisson-statistically distributed collisions with a fixed (velocity
independent) collision frequency (inverse of collision time). After
the collision the ion acquires a new velocity distributed according to
the supposed neutral distribution function, which is taken to be a
Maxwellian shifted by the same flow velocity as the external ion
drift. In this situation no external field is required to sustain the
drift, and the external ion distribution is the same shifted
Maxwellian. Its temperature is $T_i=T_e/100$, which is appropriate for
approximately room temperature ions in a partially ionized plasma. The
external drift velocity $v_d$ is in the direction $\hat{\bf z}$, and
is expressed as a Mach number by dividing the velocity by the cold-ion
sound speed $c_s=\sqrt{T_e/m_i}$: $M=v_d/c_s$.

Ions are continuously injected at the boundary at a rate and with a
distribution corresponding to the external plasma density, temperature
and flow. Ions are removed from the simulation when they leave the
computational box. Poisson's equation for the potential is solved for
each time step with charge density determined by the self-consistent
electron and ion densities, assigned by the so-called
cloud-in-cell\cite{birdsall91} algorithm to the finite difference
lattice. The code is advanced in time by a leap-frog scheme till it
converges to a steady state. The boundary conditions on Poisson's
equation at the edge of the computational mesh are designed to model a
perturbed region within a wider unperturbed background. On the leading
edge, the normal potential gradient is set to zero. On the sides of
the box, to which the external drift is tangential, the potential
gradient is set to zero in the direction $M \hat{\bf z} + \hat{\bf
  r}$, where $\hat{\bf r}$ is the direction of the cylindrical radius
$r=\sqrt{x^2+y^2}$. This oblique choice acts approximately as a
non-reflecting boundary condition for the oblique perturbations of the
wake. At the trailing face, the potential boundary condition is
equivalent to $\partial^2\phi/\partial z^2 = -\phi/(M\lambda_{De})^2$
which is non-reflecting for the dominant wavelength of
perturbation. These choices minimize the effects of the boundary
proximity on the potential solution in the inner region, and hence
allow a relatively compact computational domain without significant
boundary effects, as has been confirmed by trials with larger box.

The force on a grain is obtained by surrounding it by a sphere
across which the momentum flux of ions, electrons, and fields
(i.e.\ the Maxwell stress) is accumulated. When collisions are
present, they give rise to a bulk momentum source (or sink) within the
sphere's volume which must also be accounted for. The way in which
this is done has been explained elsewhere\cite{Patacchini2008}. Then
in steady state, the value obtained for the total momentum flux to the
grain ought to be independent of the size of the sphere --- and it is!

The calculations presented here are for nominally point grains, which
are known from previous simulations\cite{hutchinson11} to give results
very similar to finite-sized small grains. However, especially for
collisional cases included here, the code actually behaves as if a
small absorbing object were present at the point. Ions that happen to
step very close to the point charge experience an extremely large
acceleration (which is evaluated directly in terms of Coulomb force).
The impulse they acquire in a single time step can then be sufficient
to impart (unphysically) a high velocity sufficient to move them
directly out of the mesh on the next step. Thus they are removed from
the simulation, effectively as if they were absorbed when approaching
the grain too closely. For force calculations it is preferable simply
to drop them from the simulation when their single-step impulse
exceeds $5m_ic_s$, rather than accumulating their big unphysical
momentum impulse which enhances the noise level of the force
measurement. The removal of ions is thus equivalent to the presence of
an absorbing grain whose radius depends upon grain-charge and
time-step duration. For the parameters used here, this effective radius
does not exceed approximately $0.02\lambda_{De}$.


Calculations are performed in dimensionless units. The lengths
reported here are all normalized to the Debye length $\lambda_{De}$,
in figure labelling sometimes written abbreviated as $\lambda$.

\section{Single-grain Wake}

At sonic ion flow speeds, it has been shown by two-grain
simulations with COPTIC\cite{Hutchinson2011} that the transverse force
on a grain in the collisionless wake of another is well approximated
as being given by the gradient of the potential of the wake of the
upstream grain in the absence of the downstream grain, acting on
the downstream grain's charge. In other words, the force on the
downstream grain can be derived directly from the upstream
grain's unperturbed wake field. Although this identity has not
prevously been established for deeply subsonic flow speeds, it
suggests that we should first examine what happens to the single-grain
wake potential as the Mach number is decreased. Prior calculations of
the wake potential (e.g.\cite{winske00,Miloch2010}) have rarely if
ever systematically explored values less than $M=0.5$, but
linear response calculations are being pursued\cite{ludwig12}.

\begin{figure}[htp]
  \subfigure[]{\includegraphics[height=2.8in]{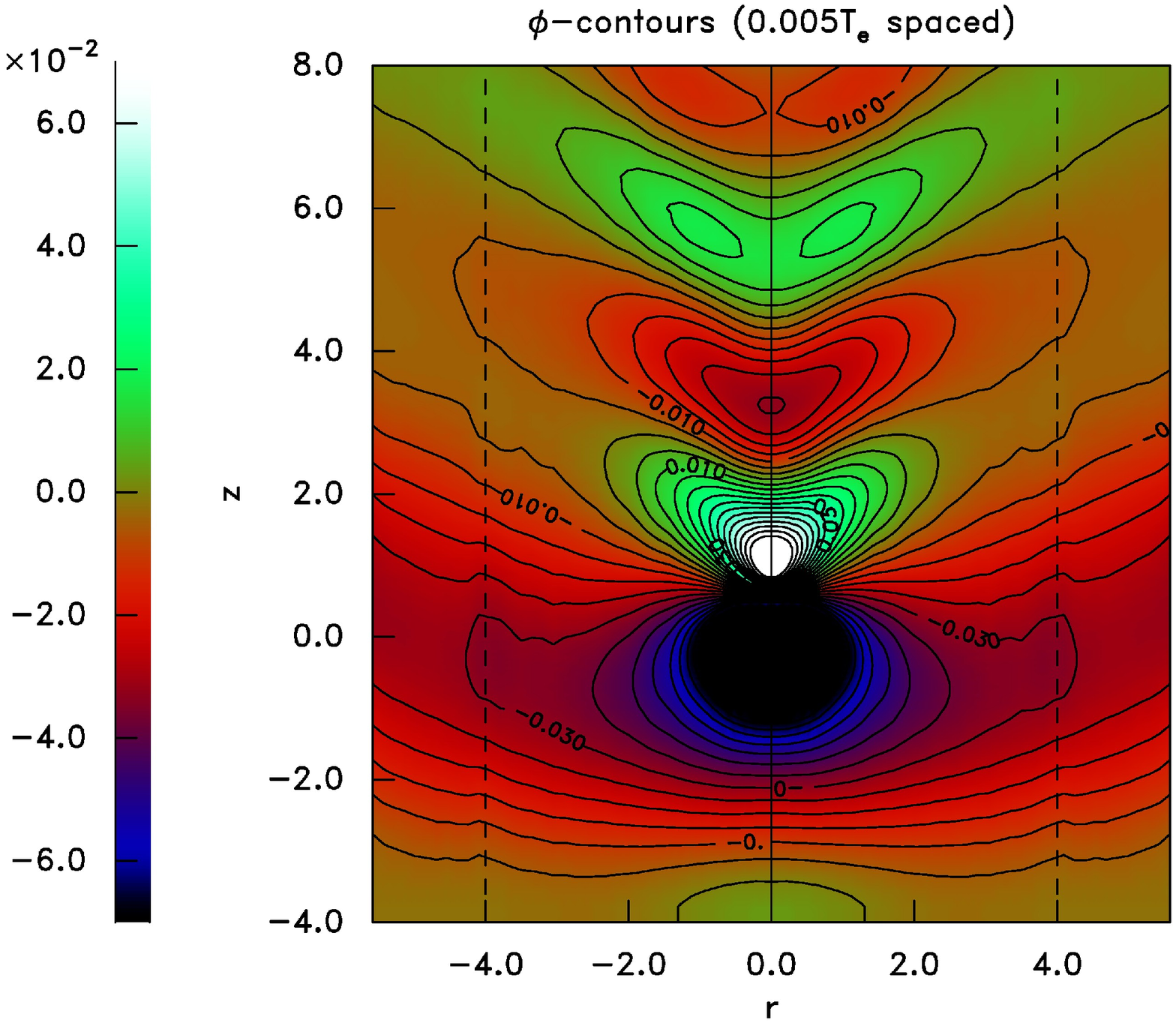}}
  \subfigure[]{\includegraphics[height=2.8in]{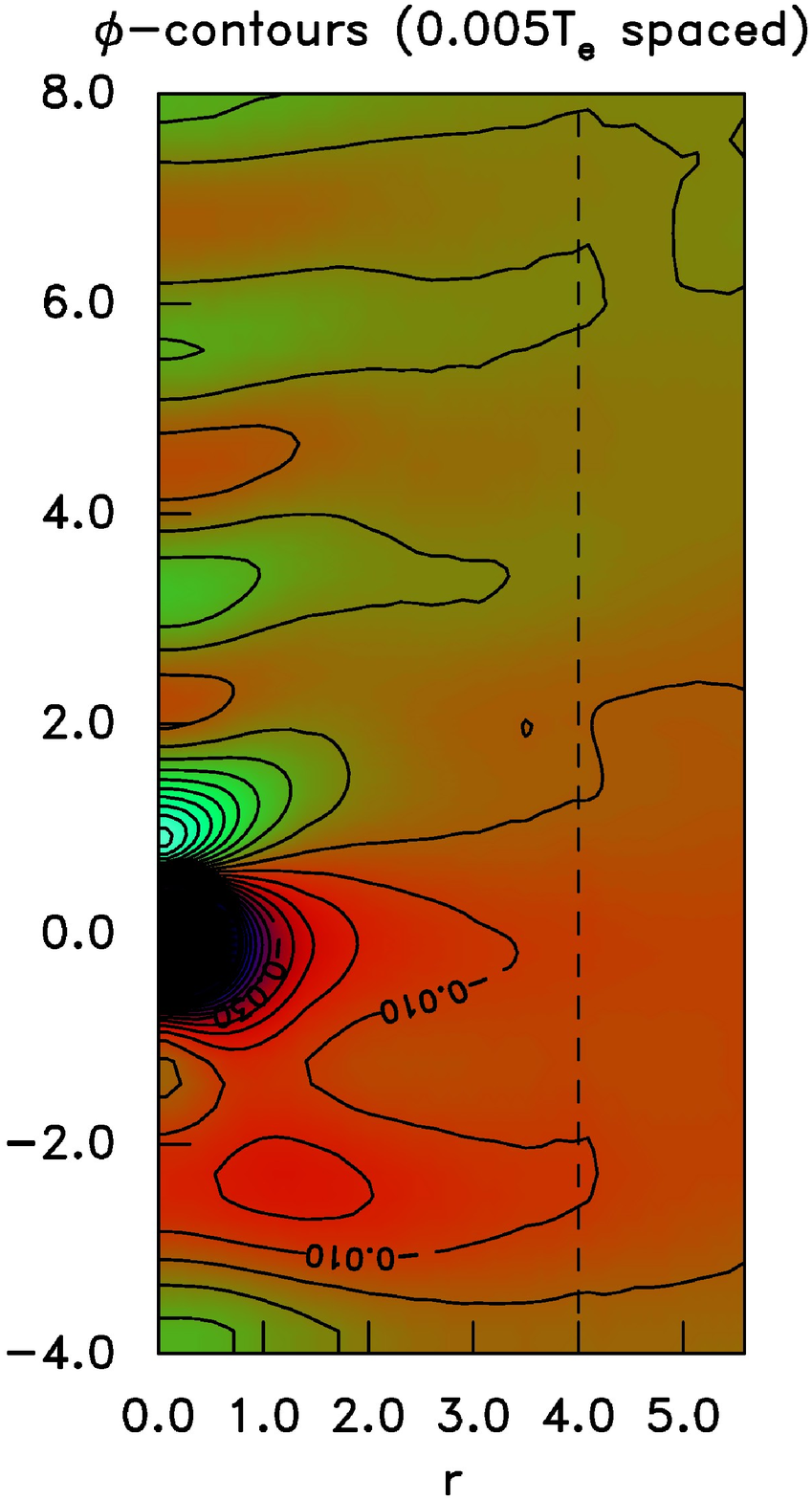}}
  \subfigure[]{\includegraphics[height=2.8in]{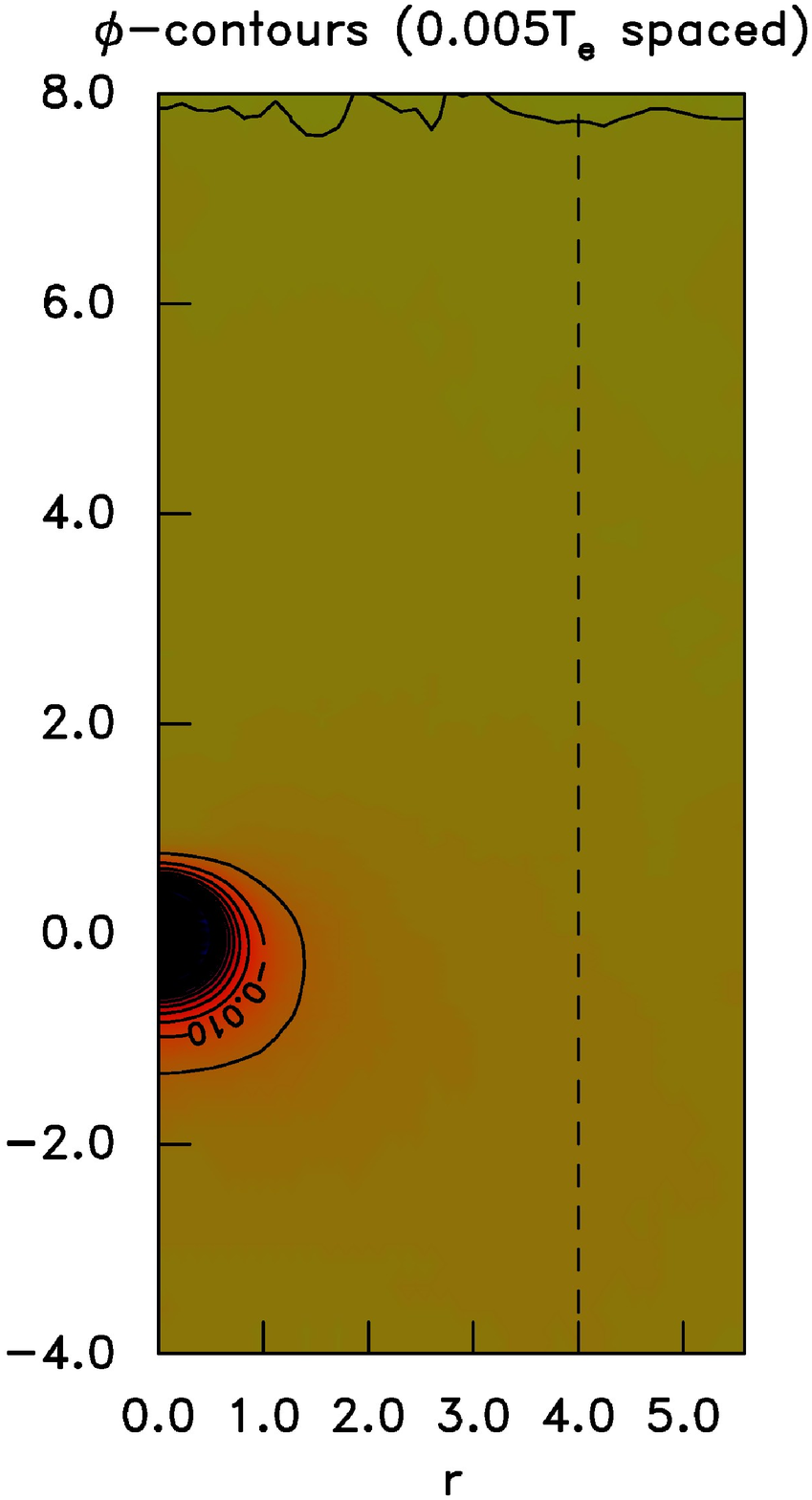}}
  \caption{Wake potential contours for single grain. (a) $M=0.6$, (b)
    $M=0.4$, (c) $M=0.2$.
    Collisionless plasma, non-linear
    regime: point charge $\bar{Q}=-0.2$. \label{phi1p2v6}}
\end{figure}

In Fig.\ \ref{phi1p2v6} are shown contour plots of the wake potential
of a single grain. These are in principle rotationally symmetric,
i.e.\ functions only of $r$ and $z$; so they are obtained by averaging
the three-dimensional solutions of COPTIC over cylindrical angle
$\theta$. (This gives a total $r$-extent up to $\sqrt{2}$ times the
box half-width, but values beyond that half-width of 4 become
increasingly subject to distortion by the boundary proximity.)
Fig.\ \ref{phi1p2v6}(a) gives the result for $M=0.6$, plotting also
the mirror image (at ``negative r'') to emphasize the conical
structure of the wake oscillations. Figs.\ \ref{phi1p2v6}(b) and (c)
show only the positive half, for $M=0.4$ and $M=0.2$ respectively,
using the same contour values. By the time the Mach number has been
lowered to 0.2 all the oscillations in the wake have vanished, and one
is left with simply the potential well of the negatively charged
grain.

\begin{figure}[htp]
  \subfigure[]{\includegraphics[width=3.2in]{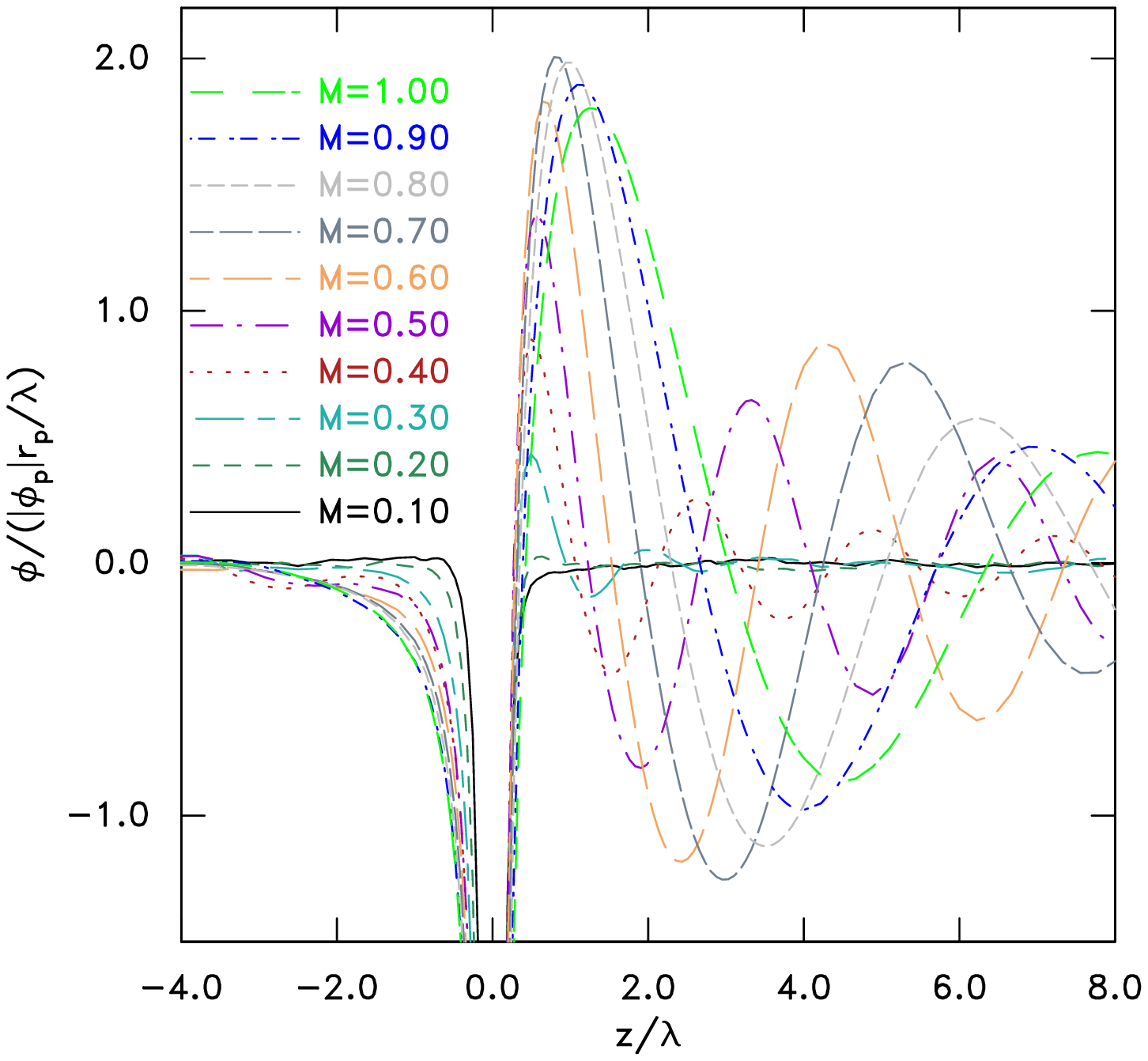}}
  \subfigure[]{\includegraphics[width=3.2in]{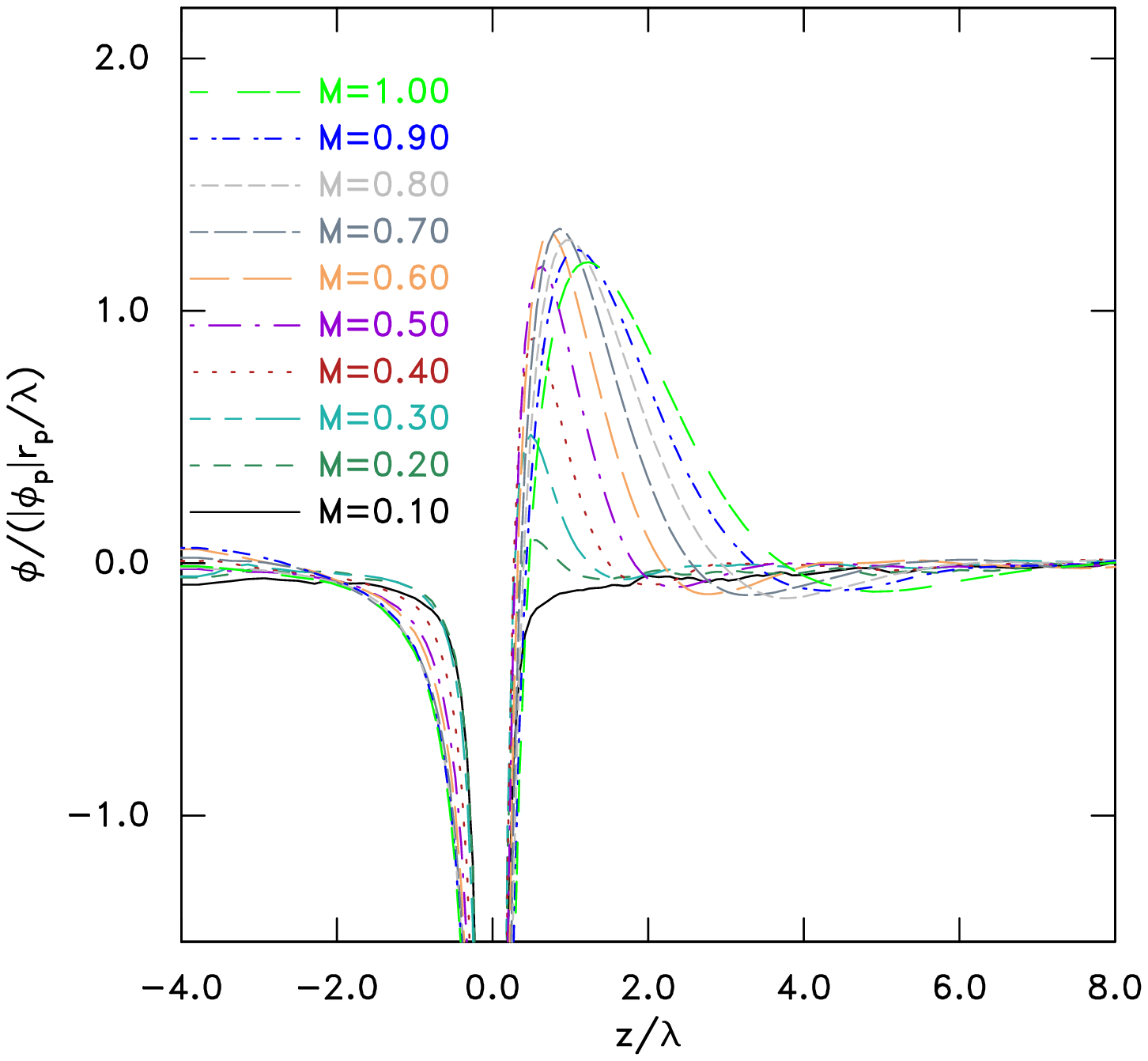}}
  \caption{Wake axial potential profile for mach numbers
    $M$. Near-linear regime: point charge
    $\bar{Q}=-0.02$. (a) Collisionless plasma, (b)
  Collisional plasma (collision time
  $1.3\lambda_{De}/c_s$). \label{phizp02}} \end{figure}

\begin{figure}[htp]
  \subfigure[]{\includegraphics[width=3.2in]{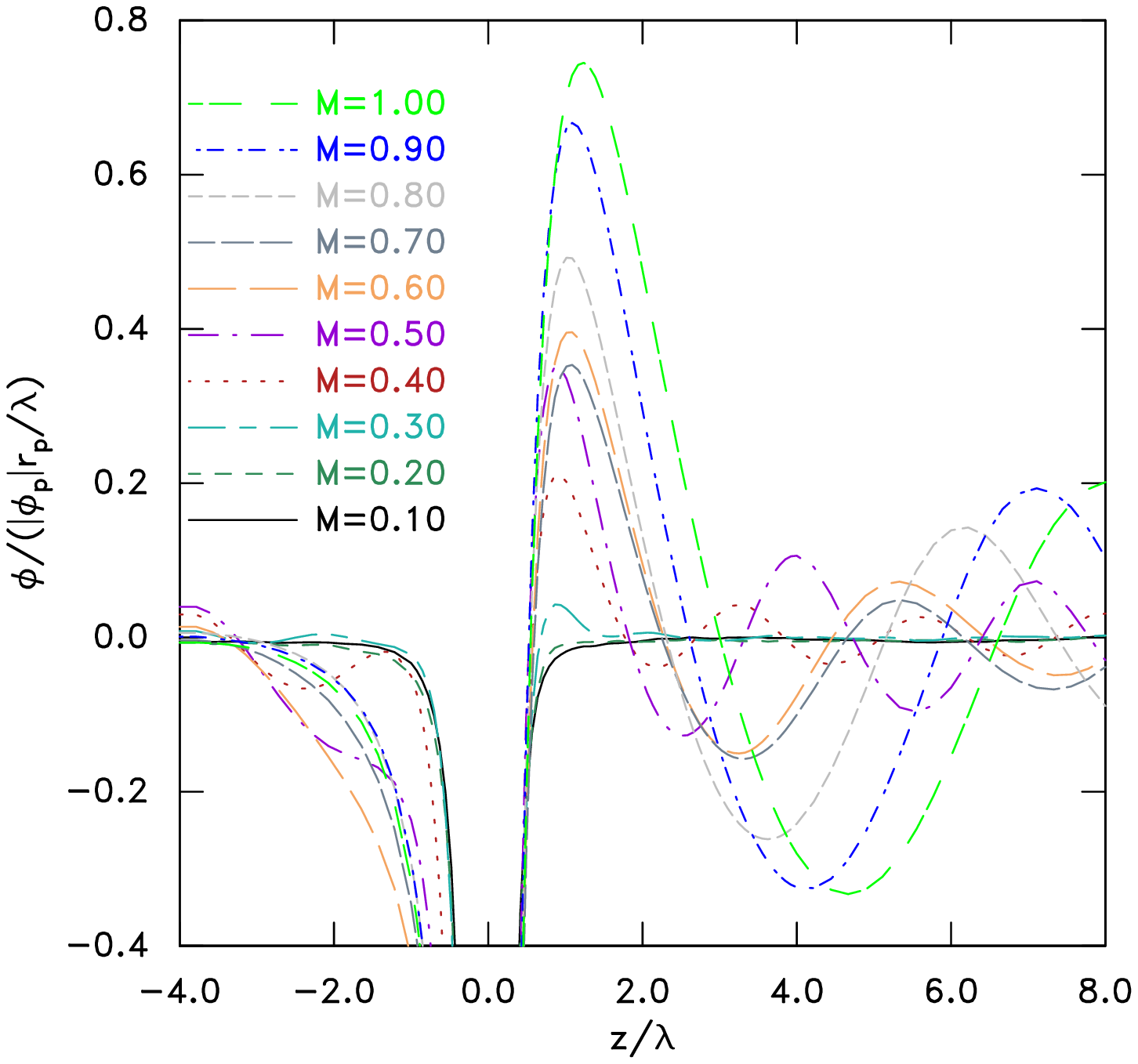}}
  \subfigure[]{\includegraphics[width=3.2in]{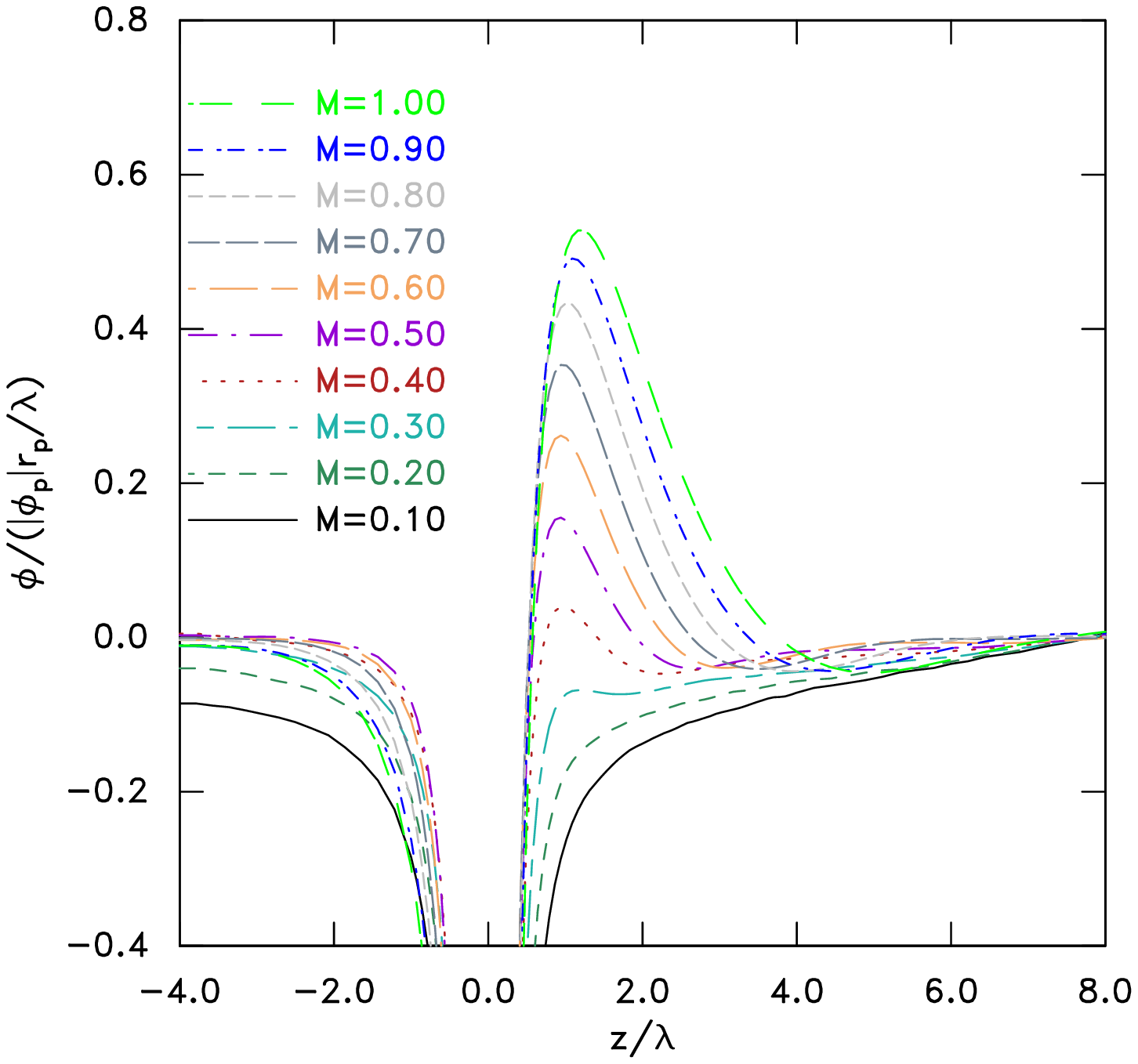}}

  \caption{Wake axial potential profile for mach numbers
    $M$. Non-linear regime: point charge $\bar{Q}=-0.2$.  (a)
    Collisionless plasma, (b) Collisional plasma (collision time
    $1.3\lambda_{De}/c_s$).\label{phizp2}}
\end{figure}

In Figs.\ \ref{phizp02} and \ref{phizp2} are shown potential profiles
along the axis of the simulation ($x=0$, $y=0$). The value of
normalized grain charge $\bar{Q}\equiv
Q/(4\pi\epsilon_0\lambda_{De}T_e/e)$ (where $Q$ is the unnormalized
grain charge) determines the strength of the interaction. For
grains small compared with $\lambda_{De}$ (the situation considered
here), $\bar{Q}\approx(e\phi_p/T_e)(r_p/\lambda_{De})$, where $\phi_p$
is the grain's surface potential. An isolated grain floats at a
surface potential of roughly $-2T_e/e$, therefore $\bar{Q}$ can be
considered to represent the size of the grain relative to the Debye
length: $r_p\approx \lambda_{De}|\bar{Q}|/2$. When the charge is
small, $\bar{Q}=-0.02$ as in Fig.\ \ref{phizp02}, prior investigations
have shown the response to be nearly linear and to agree with linear
response formalism\cite{hutchinson11}. But when it is ten times larger, $\bar{Q}=-0.2$,
as in Fig.\ \ref{phizp2} there are already strong non-linear
modifications of the response even for sonic flow. Slower flow
velocities, such as we are investigating here, are expected to
experience stronger non-linearities than prior calculations.  The
modification is predominantly a suppression of the magnitude of the
oscillatory potential in the wake. This shows up when comparing
Figs.\ \ref{phizp02} and \ref{phizp2}, by the fact that the wake
potential peak normalized to $\bar{Q}$ is roughly three times smaller
for the non-linear case than the near-linear case. (Exact linearity
would give equal normalized potential $\phi/\bar{Q}$.) We also compare
 (a) collisionless plasmas with (b) plasmas in which substantial
ion-neutral charge-exchange collisions occur. Many
experiments have sufficient background neutral pressure for collisions
to be important.
The collision time chosen here is $\tau = 1.3 \lambda_{De}/c_s$
giving rise to collision length of approximately $\lambda_{De}$ for
ions at the sound speed. This collisionality is a reasonable
approximation for Debye lengths somewhat less than a millimeter in
Argon gas of
pressure a few tens of Pa, which is typical of some experimental
conditions. (But it is not intended to model experimental collision
effects precisely since the physical collision frequency is not
independent of ion velocity.)

As the Mach number is decreased, for the collisionless plasma, the
wavelength of the potential oscillations in the collisionless wake
continues to be rather well given by the expression $2\pi \lambda_{De}
M$ established for transonic flow. This value can be understood
qualitatively as a radial compressional ion oscillation of frequency
$\omega_{pi}$ (induced by the presence of the grain) convected by the
flow at speed $M c_s$. The oscillations persist for $M\ge 0.4$ but
disappear rather abruptly for smaller drift velocity. In these plasmas
with $T_i/T_e=0.01$, the ion thermal velocity is $\sqrt{T_i/m_i}=0.1
c_s$. Therefore, disappearance of the oscillations at $M=0.3$ (and
below) occurs where linear Landau damping becomes strong,
conventionally considered to be at phase velocity three times the
thermal velocity. 

This interpretation is supported by other COPTIC calculations
(not shown here) at temperature ratio $T_i/T_e=0.1$, which show wake
oscillations disappearing below approximately $M=0.8$ (2.5 times the
thermal velocity). The trailing potential peak then disappears below
approximately $M=0.5$, suggesting that it is less susceptible to
Landau damping per se.

Substantial charge-exchange collisionality, when collision time equals
$1.3 \lambda_{De}/c_s$, has the effect of suppressing the oscillatory
wake. However, collisionality does not remove the large positive
potential peak that immediately follows the grain. It only reduces
its height by up to 40\%. That peak is attractive to other
negatively-charged grains, and is considered the cause of vertical
alignment of pairs of grains suspended in a sheath edge.

\begin{figure}[p]
  \subfigure[]{\includegraphics[width=3.2in]{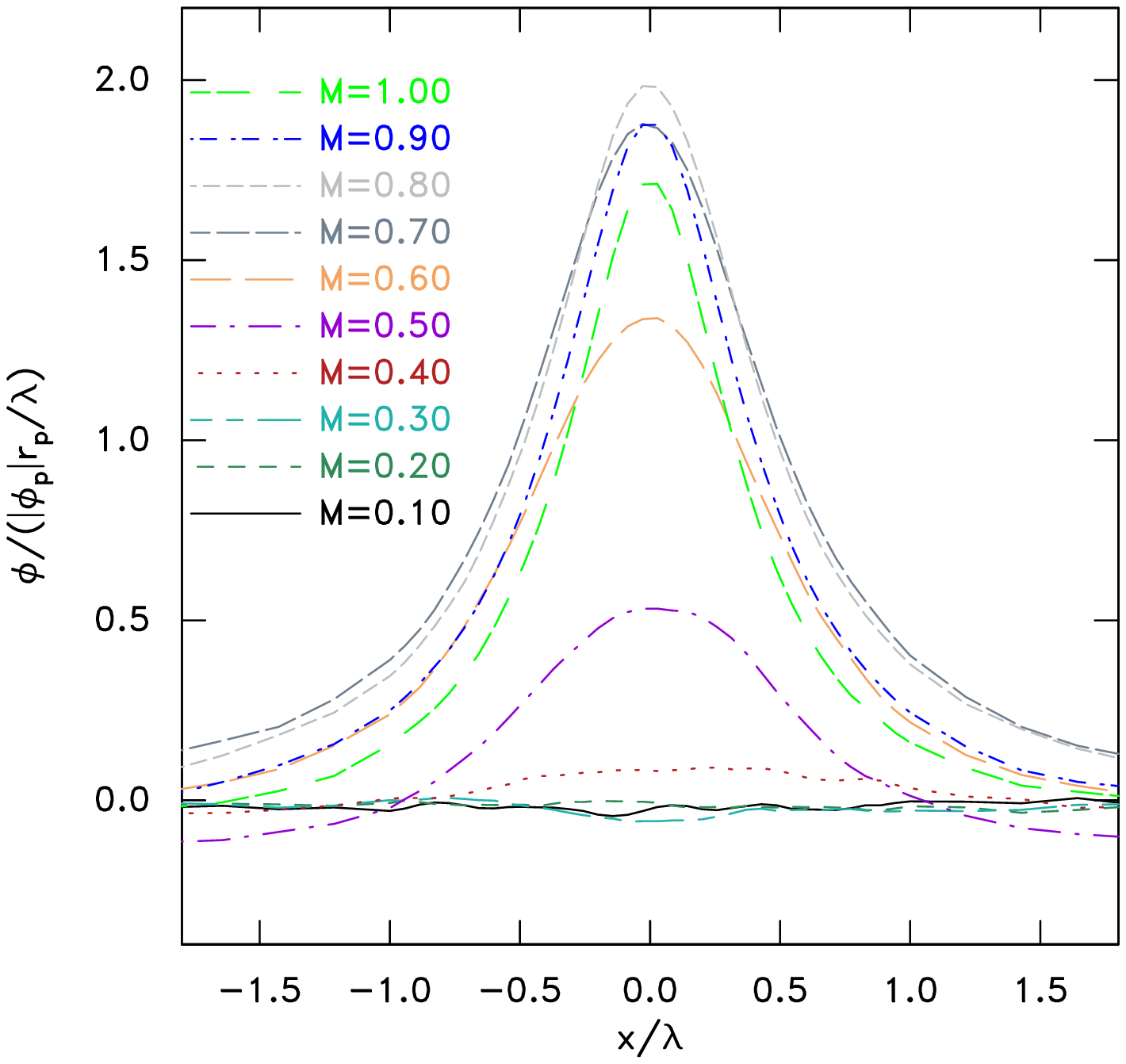}}
  \subfigure[]{\includegraphics[width=3.2in]{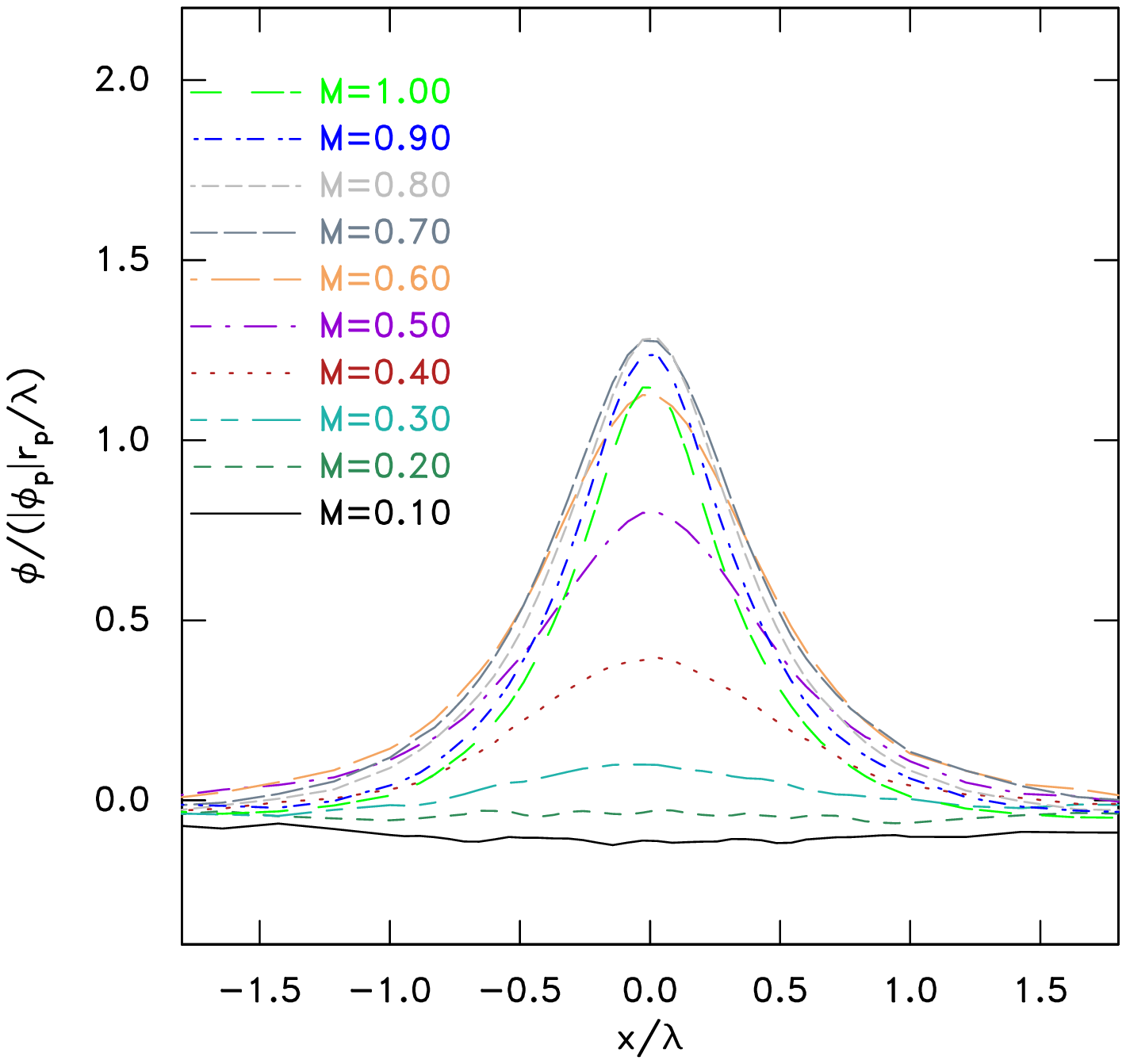}}
  \caption{Wake transverse potential profile at $y=0$, $z=1$ for mach
    numbers $M$. (a) Collisionless plasma, (b) Collisional
    plasma. Near-linear regime: point charge $\bar{Q}=-0.02$. \label{phixp02}}
\end{figure}

\begin{figure}[p]
  \subfigure[]{\includegraphics[width=3.2in]{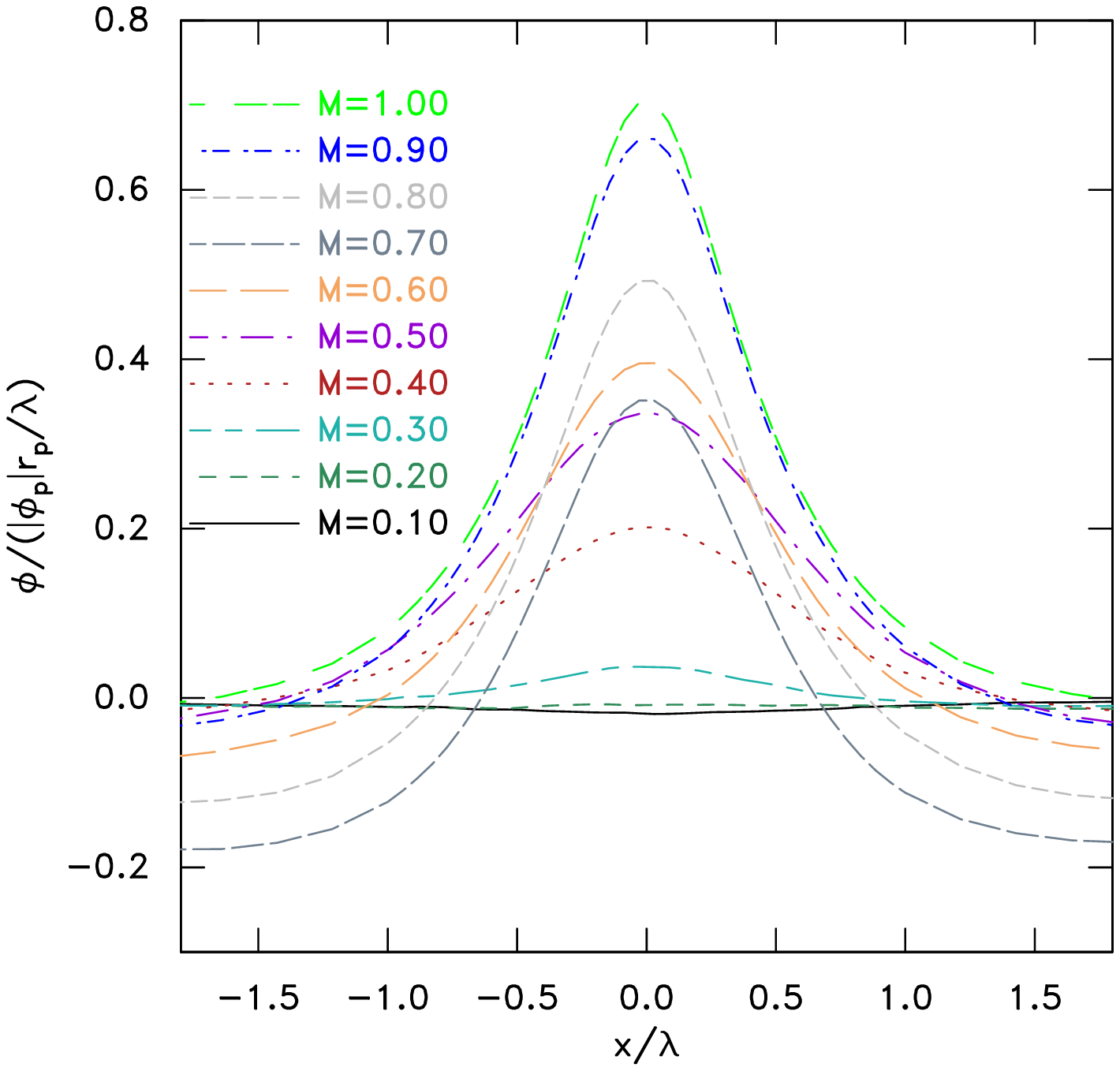}}
  \subfigure[]{\includegraphics[width=3.2in]{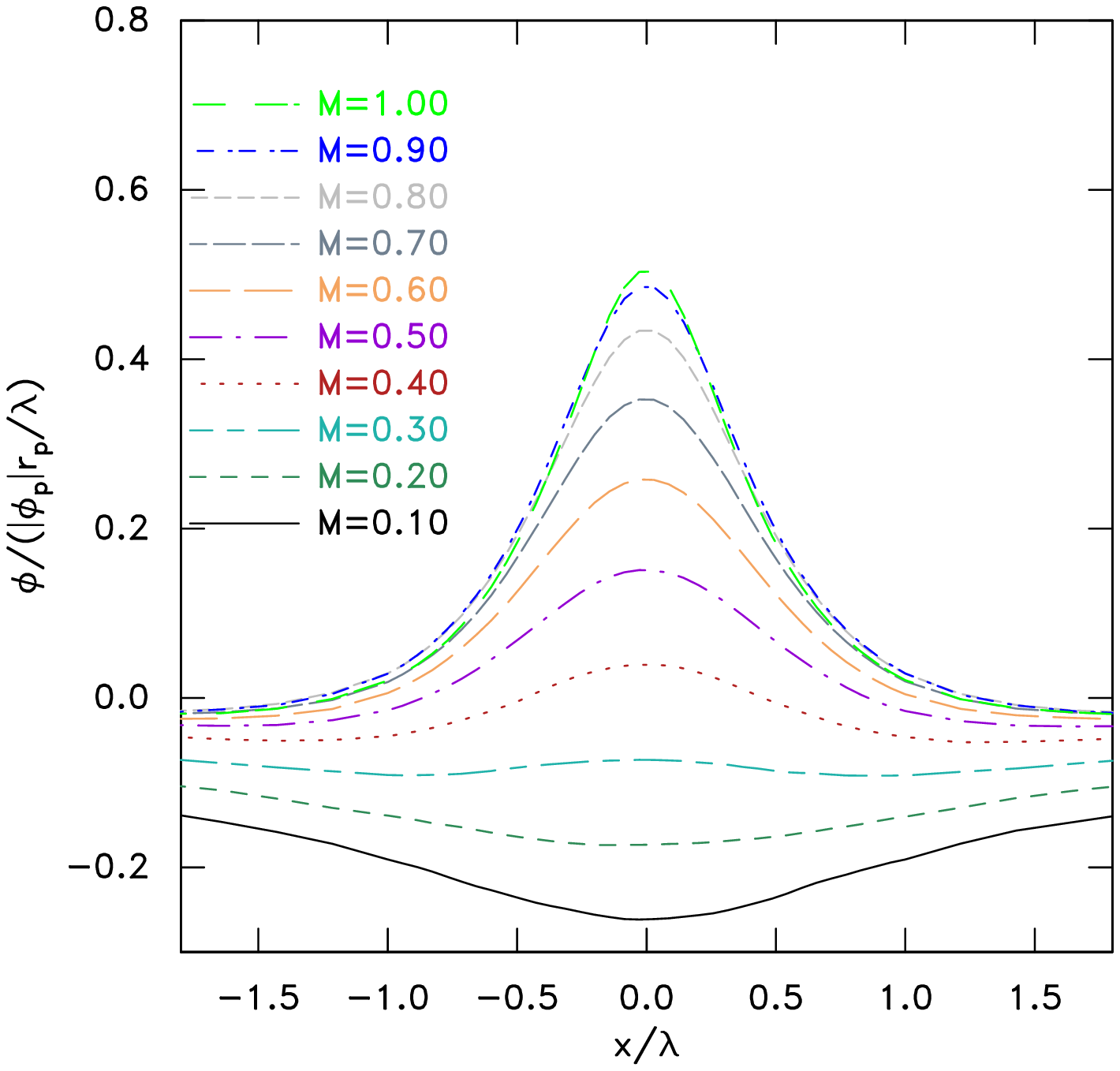}}
  \caption{Wake transverse potential profile at $y=0$, $z=1$ for mach
    numbers $M$. (a) Collisionless plasma, (b) Collisional
    plasma. Non-linear regime: point charge $\bar{Q}=-0.2$. \label{phixp2}}
\end{figure}

This trailing attractive potential region disappears for both
collisional and collisionless plasmas when $M$ decreases.  A
tiny vestige is left when $M=0.3$ which essentially vanishes for $M\le
0.2$. It is interesting to notice that this transition coincides with
the filling in of the potential depression at the \emph{upstream} side of the
grain. (Most noticeable for collisionless plasmas.) This leading
potential well has a scale-length of at least $\lambda_{De}$ for $M>0.4$, but
it fills in, yielding an upstream scale-length much shorter for
slower drifts. This is presumably a reversion from electron-shielding
to ion-shielding, the ion Debye length here being ten times shorter than
$\lambda_{De}$.

Potential profiles in the direction transverse to the flow determine
the transverse electric field, which gives a force tending to align
the grains. Examples are shown
in Figs.\ \ref{phixp02} and \ref{phixp2} at a distance downstream of one
Debye length ($z=1\lambda_{De}$, $y=0$). We observe the rapid transition from
attractive (positive) potential to practically flat potential as $M$
drops below 0.4. Actually in collisional cases, the potential becomes
slightly negative (repelling) at the lowest values of $M$.

\begin{figure}[htp]
  \includegraphics[width=5in]{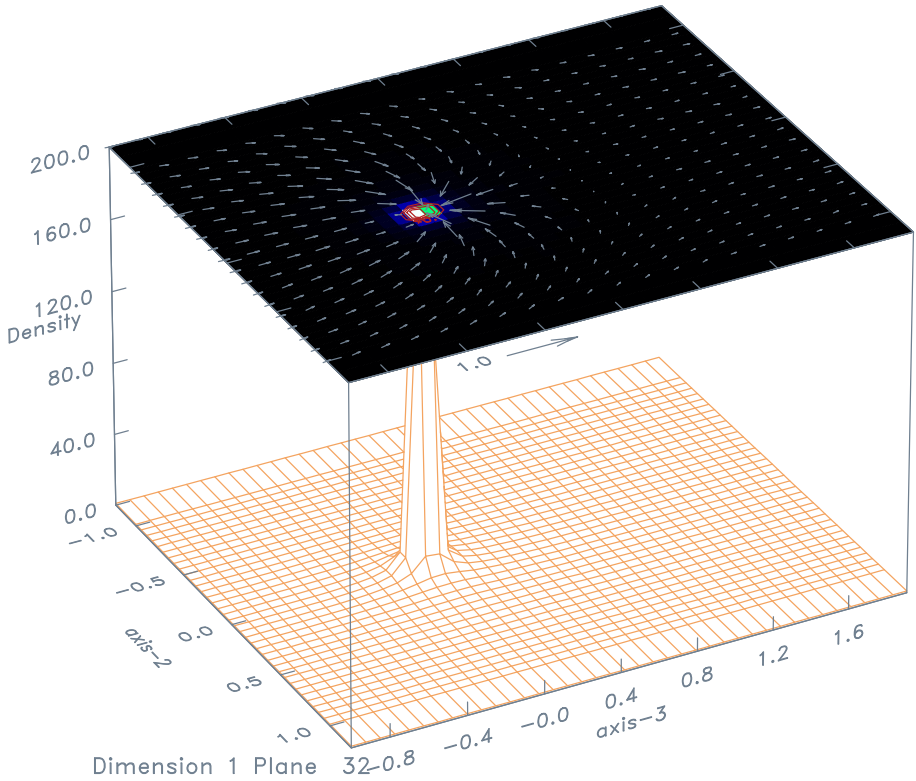}
  \caption{Flow pattern for collisional case at low drift $M=0.1$ (in
    direction ``axis-3'') in the vicinity of a large charge
    $\bar{Q}=-0.2$. Arrows indicate the magnitude and direction of ion
    mean velocity in the plane $x={\rm constant}$ through the
    origin. Ion velocity converges to compensate for the effective
    loss to the grain. The ion density (relative to
    ambient) in this plane is shown as a 3-D surface and as contours. It becomes
    extremely large in the potential well because of trapped ions: as
    much as 200 times ambient.\label{arrowv01}}
\end{figure}

The reason for the presence of the extended potential well in
collisional plasmas, most obvious in the non-linear case
(Figs.\ \ref{phizp2}(b) and \ref{phixp2}(b)), is that the effective
absorption of ions at the grain causes an inflow of ions. Its
magnitude is controlled effectively by the collision rate and
prescribed large grain-charge (and not by the somewhat ill-defined
effective computational grain size). This inflow, which is illustrated
in Fig.\ \ref{arrowv01}, becomes the dominant effect on potential
profile at low-$M$. Collisional trapping causes an enormous rise of
the ion density in the grain's potential well: as much as a factor of
200. The physics is thus making a transition at low Mach number
to the spherically symmetric collisional collection
studied in references \cite{lampe03,Hutchinson2007a}.

Summarizing the single-grain wake potential observations: for
collisionless and collisional, near-linear and non-linear plasmas, at
$T_e/T_i=100$, the attractive wake potential peak disappears (or at
least is greatly reduced) for flow Mach numbers of about 0.3 and
below.

\section{Two-grain simulations}

The full non-linear calculation of the force on the downstream
grain requires a 3-dimensional calculation with two grains. One
is at the origin, and the second having the same charge for these
calculations is placed at $z=1$, $x=0.5$, $y=0$ (in units of
$\lambda_{De}$). Comparison with Figs.\ \ref{phixp02} and \ref{phixp2}
shows that this transverse position is near the maximum of the
potential gradient, thus giving a representative measure of the
transverse, grain-aligning, force.

\begin{figure}[htp]
  \includegraphics[width=5.2in]{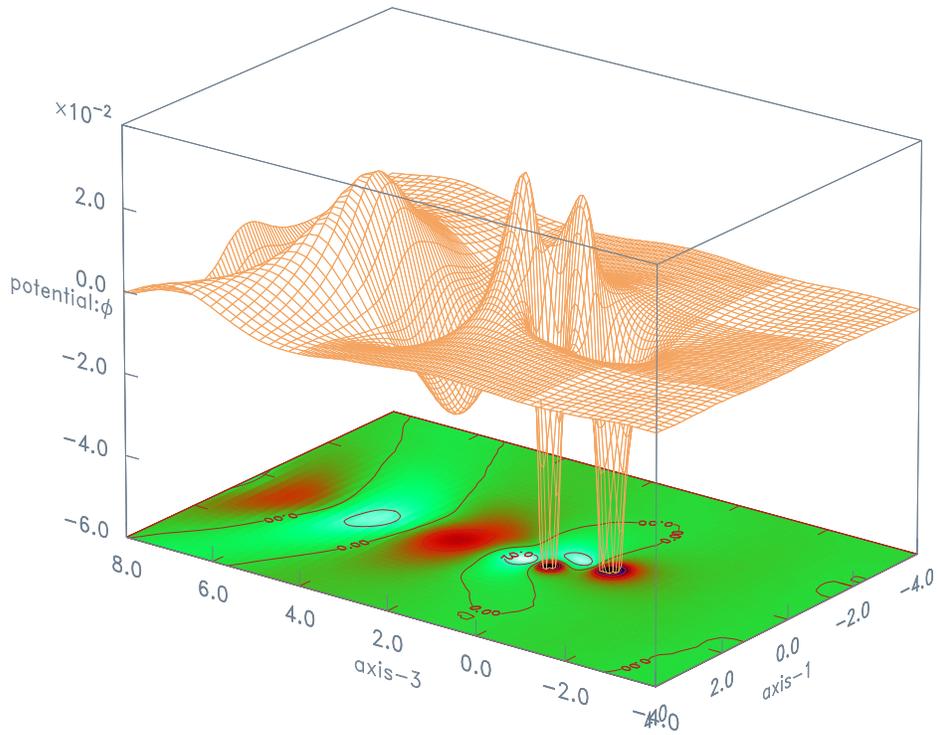}
  \caption{Wake potential for two grains at $M=0.6$. The slice
    shown is on the plane $y=0$.  Collisionless plasma, near-linear
    regime: point charge $\bar{Q}=-0.02$. \label{phip02v6}}
\end{figure}
In Fig.\ \ref{phip02v6} is shown the wake potential
calculated by COPTIC on a domain of size
$(\pm4,\pm4,-4\to+8)\lambda_{De}$ spanned by a mesh
$64\times64\times100$.
The plasma drift velocity is in the $z$-direction (along ``axis-3'').
The deep potential wells caused by the two charged grains are cut
off conveniently for viewing. Immediately behind them are two
potential peaks which are partly joined. For this $M=0.6$ case, a combined
oscillatory wake is present beyond the two peaks, which does not look
much different from a typical one-grain wake.

\begin{figure}[htp]
  \includegraphics[width=4in]{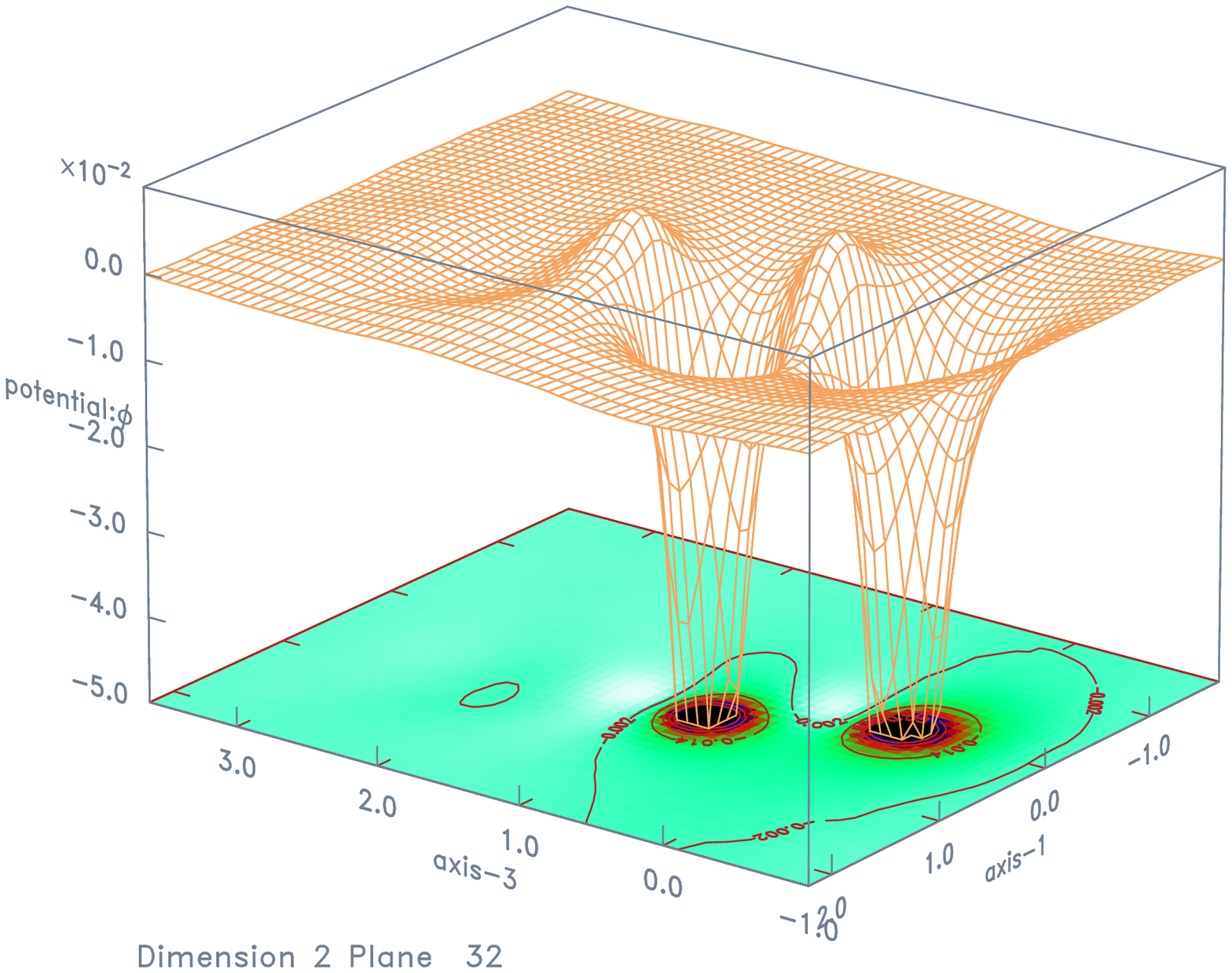}
  \caption{Wake potential for $M=0.3$. The slice shown is on the plane $y=0$.
    Collisionless plasma, near-linear
    regime: point charge  $\bar{Q}=-0.02$. \label{phip02v3}}
\end{figure}

\begin{figure}[htp]
  \includegraphics[width=4in]{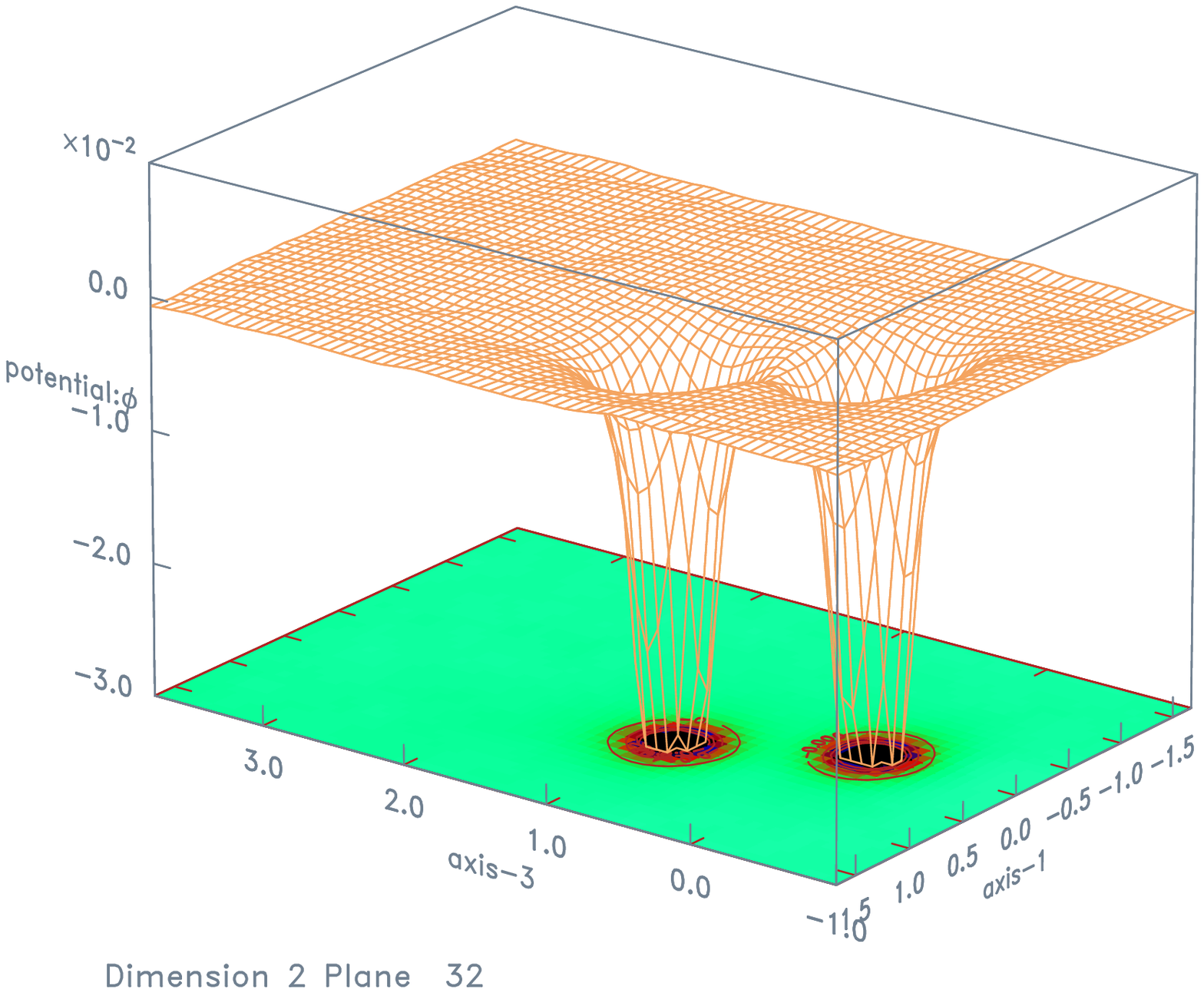}
  \caption{Wake potential for $M=0.1$. The slice shown is on the plane $y=0$.
    Collisionless plasma, near-linear
    regime: point charge  $\bar{Q}=-0.02$. \label{phip02v1}}
\end{figure}
Fig.\ \ref{phip02v3} shows the case $M=0.3$, where the oscillatory
wake damps out very quicky. Therefore the plot is focussed on the
inner region.  (The full simulation domain is the same as in
Fig.\ \ref{phip02v6} for all calculations in this paper.) The result
shows only the vestiges of the trailing positive potential peaks and
negligible oscillations.
Fig.\ \ref{phip02v1} is for $M=0.1$ where even those vestiges have
disappeared, and the only relevant remaining features are the two
potential wells of the two grains.

\begin{figure}[htp]
  \subfigure[]{\includegraphics[height=3.in]{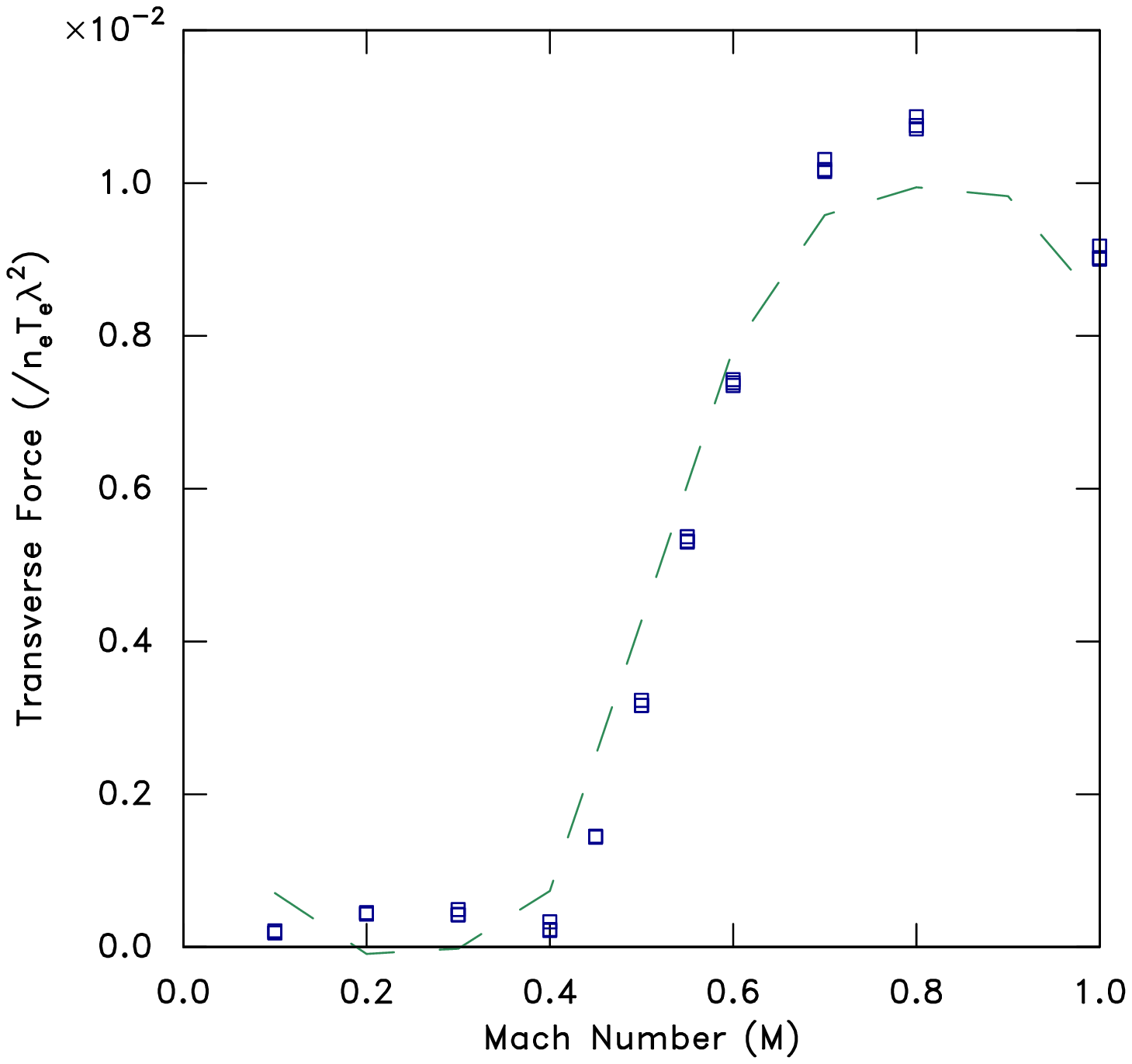}}
  \subfigure[]{\includegraphics[height=3.in]{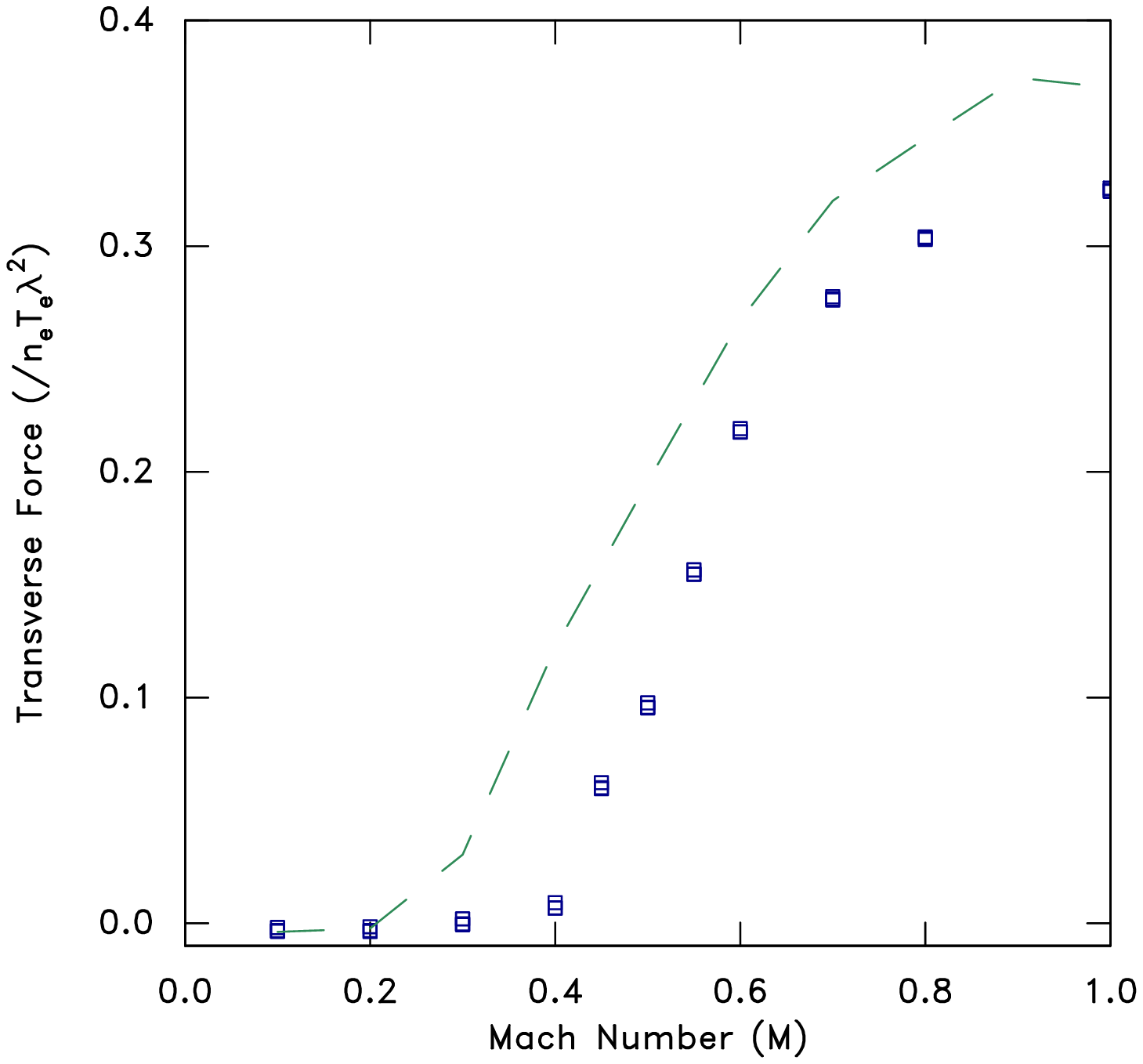}}
  \caption{Transverse force in collisionless plasma on an equal
    grain at downstream position $z=1$, $x=0.5$, $y=0.$ versus flow
    mach number $M$. Points: two-grain simulation, total force;
    dashed line: gradient of one-grain wake potential.  (a)
    near-linear regime: point charge $\bar{Q}=-0.02$, (b) non-linear
    regime point charge $\bar{Q}=-0.2$.\label{forcevsMp02}}
\end{figure}

A large number of simulation ions (about 100 million) in the code is
required to obtain the \emph{force} on the grains without excessive
noise. Each Mach number requires a different simulation, of course,
each of which is a moderately expensive calculation (about 1 hour of
128 processors for 1000 time steps). The force on the downstream grain
is measured in the code by integration of the total momentum flux
across each of three different spheres (of radius
0.3, 0.4, and 0.5 $\lambda_{De}$) centered on the downstream grain. A
perfect steady state calculation ought to give the same total force
for each of these spheres. In Fig.\ \ref{forcevsMp02} these different
measures are all three plotted to give an indication of the (rather
small) uncertainty arising mostly from mesh resolution. The good
reproducibility of the small forces was also documented by performing
equivalent runs but for different choice of coordinates: $x=0$,
$y=0.5$ or $x=-0.5$, $y=0$, etc.

Collisionless results for the transverse force are shown in
Fig.\ \ref{forcevsMp02} for two different grain-charge values. 
Attractive force, towards the $z$-axis, is plotted positive. In other words,
the ordinate is minus the $x$-force when the $x$-position is positive.
The
two-grain simulation results (points) are compared with the force
on the downstream grain that would arise from the transverse
electric field that exists in the wake of the upstream grain alone (line).
It has been shown previously that at sonic speeds the single-grain
wake field-force agrees quite well with the full two-grain force
calculation. This agreement is observed here also at low Mach
number. For grain-charge small enough to be in the near-linear regime
(Fig.\ \ref{forcevsMp02}(a)) the agreement is very good. In the nonlinear
regime (Fig.\ \ref{forcevsMp02}(b)) it is not as good, especially in the
transition region near $M=0.4$. The two-grain calculation shows the
force going abruptly to near zero, whereas the transverse wake field
gradient reaches near zero only at a lower Mach number $M\sim 0.3$.

\begin{figure}[htp]
  \subfigure[]{\includegraphics[height=3.in]{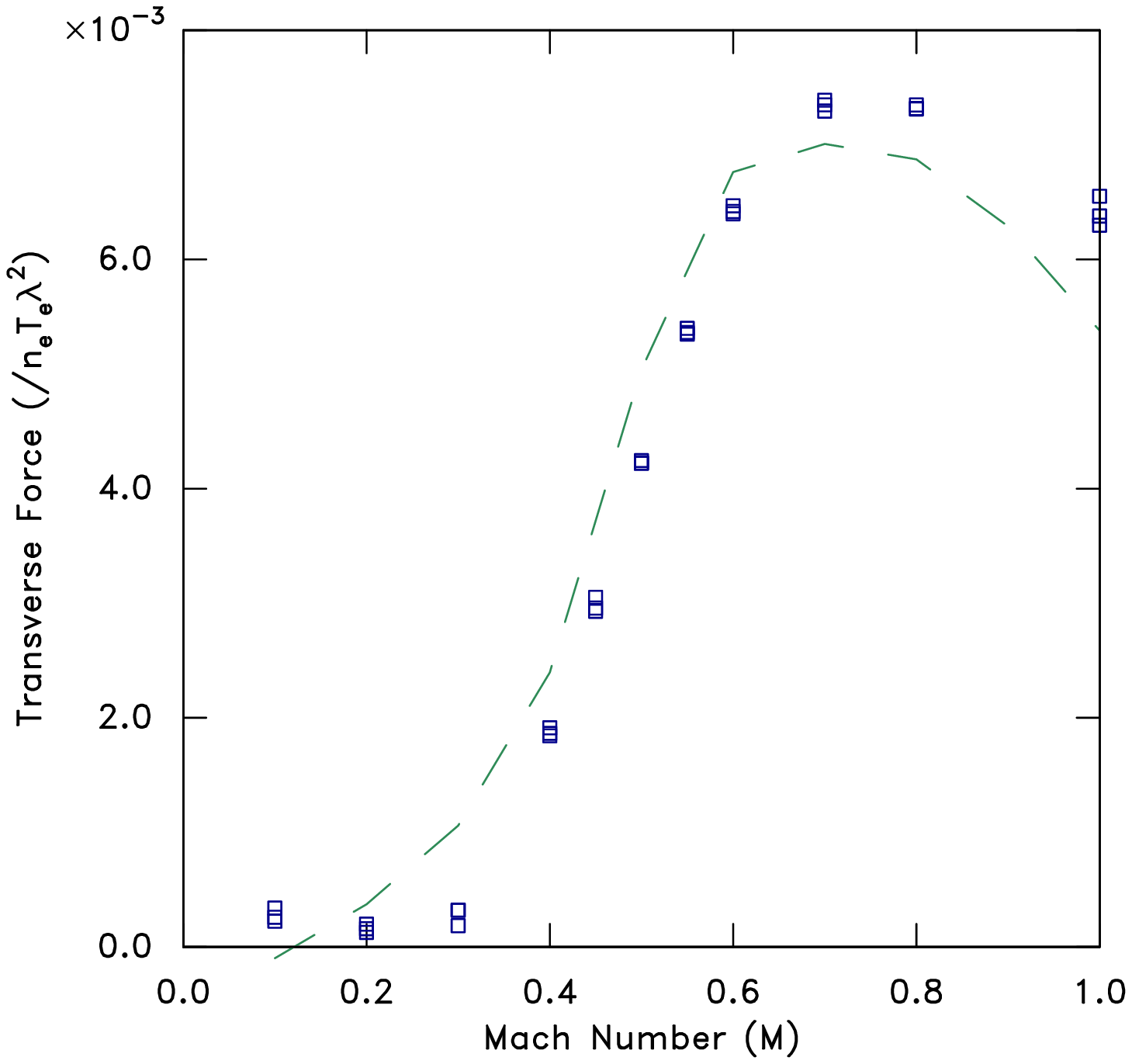}}
  \subfigure[]{\includegraphics[height=3.in]{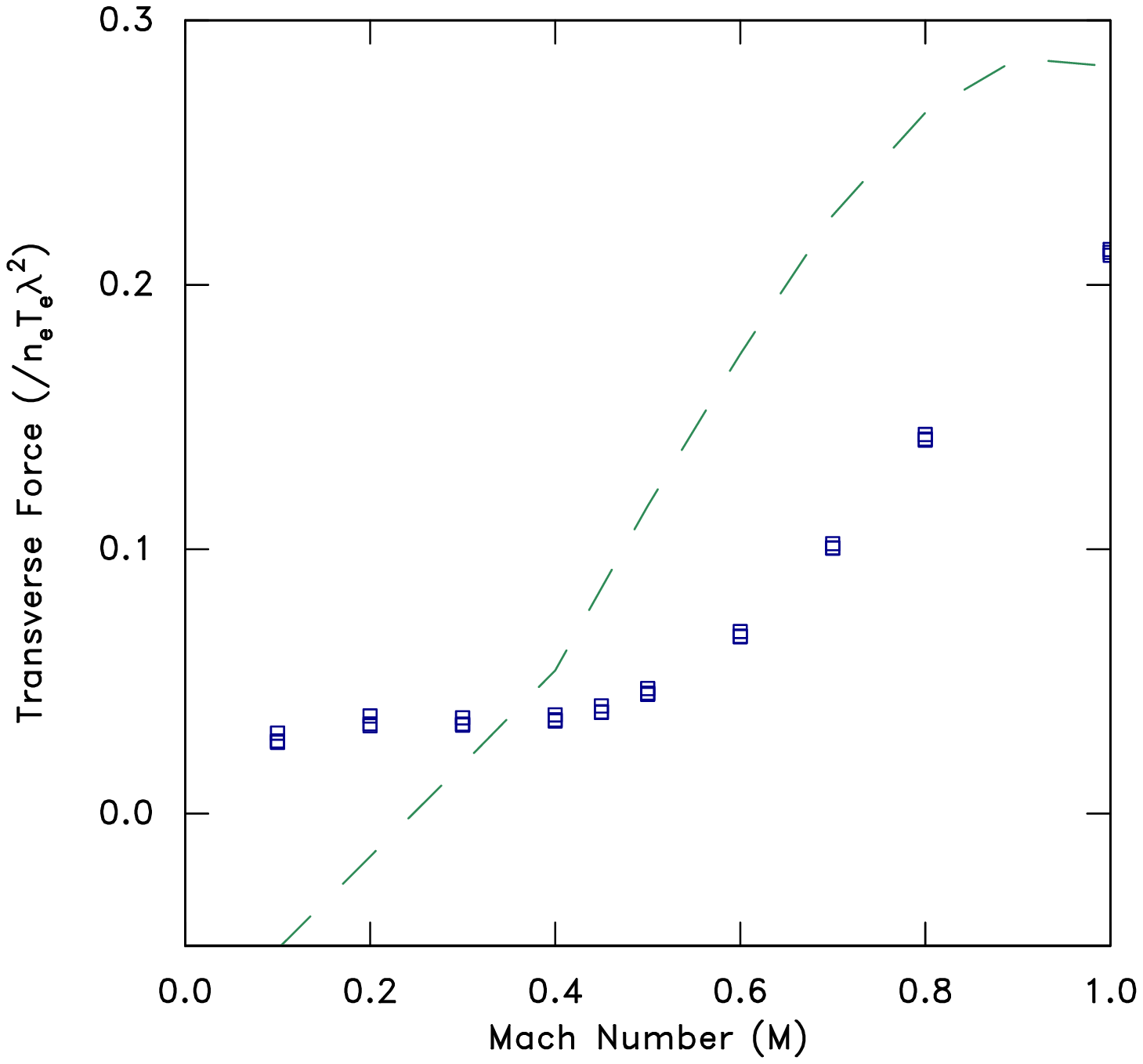}}
  \caption{Transverse force in collisional plasma on an equal grain
    at downstream position $z=1$, $x=0.5$, $y=0.$ versus flow mach
    number $M$. Points: two-grain simulation, total force; dashed
    line: gradient of one-grain wake potential.  (a) near-linear
    regime: point charge $\bar{Q}=-0.02$, (b) non-linear regime point
    charge $\bar{Q}=-0.2$.\label{forcevsMp02c}}
\end{figure}

When neutral collisions are present, the transverse force on the
downstream grain is modified as illustrated in
Fig.\ \ref{forcevsMp02c}. In the near-linear charge-magnitude regime
(Fig.\ \ref{forcevsMp02c}(a)) there is still quite good agreement
between the full two-grain simulation force and the single-grain wake
field force. The overall magnitude of the force is suppressed by about
30\% from the collisionless case, and it approaches zero for
$M\lesim0.35$. However, in the non-linear case, there are major
discrepancies between the actual transverse force and the wake-field
force. The collisions suppress the force at intermediate Mach numbers
($0.5\le M \le 1$) much more than the single-grain wake field. And at
low Mach numbers the force remains attractive (plotted positive here)
whereas the wake-field actually reverses in sign below $M=0.3$.  The
residual attractive force appears to arise because the ion drag
perturbation on the downstream particle becomes more important than
the potential gradient. The convergence of the transverse velocity,
illustrated in Fig.\ \ref{arrowv01}, then overcomes the electric
repulsion of Fig.\ \ref{phixp2}(b)). We cannot rule out, however, that
there are other synergistic two-particle effects that contribute.

The residual force might have significance for the experiments that
observe grain alignment even at low flow
velocity\cite{arp12}. Fig.\ \ref{forcevsMp02c}(b) shows non-zero attractive
(transverse aligning) force in the non-linear collisional case even at
$M=0.1$. The force is quite a small fraction (perhaps 10\%) of the
aligning force at $M=1$, but might be large enough to be significant.
The magnitude of charge in this case is approximately what would be
acquired by a grain of radius $0.1\lambda_{De}$, which is probably
rather larger than most experiments. Ten times smaller grains
(Fig.\ \ref{forcevsMp02c}(a)) appear to have little residual
attraction at low $M$.

\section{Conclusion}

It has been shown that the oscillations in the wake of a small charged
grain disappear because of damping: either collisional damping when
the flow velocity is large, or Landau damping when the flow velocity
is reduced below a few times the ion thermal velocity.  However,
neutral collisions sufficient to remove oscillations do not of
themselves remove the ion density enhancement immediately behind a
negatively charged grain, caused by focussing. Extensive computations
show that the wake density enhancement and the related potential peak
disappears quite abruptly as the flow velocity is reduced below
approximately 0.3$c_s$ (for $T_e/T_i=100$). This disappearance causes
the transverse, grain-aligning, force on a grain in the wake of
another to become negligible. The two-grain, three-dimensional,
computations confirm that for collisionless plasmas the tranverse
force is well represented by the gradient of the single-grain wake
potential for small-grains even at low Mach numbers in the near-linear
regime. It is somewhat less well represented in the non-linear
regime. For collisional plasmas in the non-linear regime the
wake-potential gradient does not represent the transverse force well,
and a small residual grain-aligning force (about one tenth of the
force for sonic flows) remains at Mach numbers down to 0.1. Although
the presence of this residual force does not allow one immediately to
rule out explanation of grain alignment and chains at low Mach
numbers, its smallness, and its absence in the near-linear regime,
suggests that other mechanisms, omitted from the present calculations,
ought to be considered. These might include forces arising from direct
neutral gas effects on the grain: drag or thermophoretic forces. They
might also include the collective effects of the other nearby grains.

\subsection*{Acknowledgements}
I am grateful to Christian Bernt Haakonsen for illuminating conversations
and to Patrick Ludwig for related collaboration. Work partially
supported by NSF/DOE Grant DE-FG02-06ER54982.

\bibliography{wake}

\end{document}